\documentclass[twocolumn,showpacs,amsmath,aps]{revtex4}
\usepackage{graphicx}
\usepackage{dcolumn}
\begin{document}
\def\eq#1{Eq.\hspace{1mm}(\ref{#1})}
\def\fig#1{Fig.\hspace{1mm}\ref{#1}}
\def\tab#1{Tab.\hspace{1mm}\ref{#1}}
\title{
Non-BCS thermodynamic properties of ${\rm H_{2}S}$ superconductor
}
\author{Artur P. Durajski$^{(1)}$}
\email{adurajski@wip.pcz.pl} 
\author{Rados{\l}aw Szcz\c{e}{\'s}niak$^{(1,2)}$}
\email{szczesni@wip.pcz.pl} 
\author{Yinwei Li$^{(3)}$}
\email{yinwei_li@jsnu.edu.cn} 
\affiliation{1. Institute of Physics, Cz{\c{e}}stochowa University of Technology, Ave. Armii Krajowej 19, 42-200 Cz{\c{e}}stochowa, Poland}
\affiliation{2. Institute of Physics, Jan D{\l}ugosz University, Ave. Armii Krajowej 13/15, 42-200 Cz{\c{e}}stochowa, Poland}
\affiliation{3. School of Physics and Electronic Engineering, Jiangsu Normal University, Xuzhou 221116, People's Republic of China}
\date{\today}
\begin{abstract}
The present paper determines the thermodynamic properties of the superconducting state in the ${\rm H_{2}S}$ compound. The values of the pressure from $130$ GPa to $180$ GPa were taken into consideration. The calculations were performed in the framework of the Eliashberg formalism. In the first step, the experimental course of the dependence of the critical temperature on the pressure was reproduced: $T_{C}\in\left<31,88\right>$ K, whereas the Coulomb pseudopotential equal to $0.15$ was adopted. Next, the following quantities were calculated: the order parameter at the temperature of zero Kelvin ($\Delta\left(0\right)$), the specific heat jump at the critical temperature ($\Delta C\left(T_{C}\right)\equiv C^{S}\left(T_{C}\right)-C^{N}\left(T_{C}\right)$), and the thermodynamic critical field ($H_{C}\left(0\right)$). It was found that the values of the dimensionless ratios: $R_{\Delta}\equiv 2\Delta(0)/k_{B}T_{C}$, $R_{C}\equiv{\Delta C\left(T_{C}\right)}/{C^{N}\left(T_{C}\right)}$, and $R_{H}\equiv{T_{C}C^{N}\left(T_{C}\right)}/{H_{C}^{2}\left(0\right)}$ deviate from the predictions of the BCS theory: $R_{\Delta}\in\left<3.64,4.16\right>$, $R_{C}\in\left<1.59,2.24\right>$, and $R_{H}\in\left<0.144, 0.163\right>$. Generalizing the results on the whole family of the ${\rm H_{n}S}$-type compounds, it was shown that the maximum value of the critical temperature can be equal to $\sim 290$ K, while $R_{\Delta}$, $R_{C}$ and $R_{H}$ adopt the following values: $6.53$, $3.99$, and $0.093$, respectively.
\end{abstract}
\maketitle
%

%
%

The experimental results, which prove that the compound ${\rm H_{2}S}$ under the influence of the high pressure ($p$) has the extremely high values of the critical temperature ($T_{C}$), were presented in December, 2014 \cite{Drozdov2014A}. In particular, it was shown that in the range of the pressures from $115$ GPa to $200$ GPa, the critical temperature increases from $31$ K to $150$ K. 
Additionally, we should underline the fact that the strong isotope effect (${\rm D_{2}S}$) was observed, which clearly suggests the electron-phonon origin of the superconducting state \cite{HirschH2S, BernsteinH2S, PapaconstantopoulosH2S, Flores-LivasH2S, AkashiH2S, ZhangH2S}. 
Interestingly, as the result of the dissociation of the starting compound, most likely of the scheme: ${\rm 3H_{2}S\rightarrow 2H_{3}S+S}$ \cite{Duan2014A}, \cite{Duan2014B}, the superconducting state with the critical temperature of up to $\sim 190$ K ($p>150$ GPa) was induced \cite{Drozdov2014A}. From the physical point of view, the obtained result indicates that the superconductor of the highest known value of $T_{C}$ was just discovered. 

Even before the release of the experimental results, the extensive theoretical studies of the superconducting condensate in the ${\rm H_{2}S}$ compound were performed in the paper \cite{Li2014A}. In the framework of the {\it ab initio} calculations, it was found that the analyzed system enters the metallic state above the pressure of $96$ GPa. Next, the existence of the superconducting state was proven in the pressures range from $130$ GPa to $180$ GPa, wherein the highest value of $T_{C}$ equal to $\sim 80$ K was obtained for $p=160$ GPa (the Cmca structure). 
It should be noted that the predictions included in the publication \cite{Li2014A} agree with the experimental results \cite{Drozdov2014A}. However, there was no structural transition observed between the phases P-1 and Cmca.
 
The results presented in the paper \cite{Drozdov2014A} prove that depending on the method to handle the expected final values of the temperature and the pressure one can prepare the compound ${\rm H_{2}S}$ or the system ${\rm H_{3}S+S}$. Note that the above results are in the agreement with the theoretical predictions that suggest the stability of the ${\rm H_{2}S}$ system below $50$ GPa \cite{Duan2014B}. On the other hand, the compound ${\rm H_{3}S}$ seems to be stable for $p\in\left<50, 300\right>$ GPa, wherein ${\rm H_{4}S}$, ${\rm H_{5}S}$, and ${\rm H_{6}S}$ are unstable in the considered range of the pressures.

The superconductor ${\rm H_{2}S}$ can be further enriched with hydrogen \cite{Strobel2011A}. The case $\rm(H_{2}S)_{2}H_{2}$ was very carefully analyzed in the work \cite{Duan2014A}. On the basis of the {\it ab initio} calculations, it was shown that the metallization of this system takes place for $111$ GPa, while for $p=200$ GPa the record value of $T_{C}$ equal to $\sim 200$ K was obtained (the $Im$-$3m$ structure).
Note that the similarly high values of the critical temperature can be obtained in the hydrogen-rich compounds of the type: ${\rm CaH_{6}}$ ($T_{C}\sim 240$ K for $p=150$ GPa) \cite{Wang2012A}, \cite{Szczesniak2013H}, ${\rm Si_{2}H_{6}}$ ($T_{C}\sim 174$ K for $p=275$ GPa) \cite{Jin2010A}, \cite{FloresLivas2012A}, \cite{Szczesniak2013C}, ${\rm B_{2}H_{6}}$ ($T_{C}\sim 147$ K for $p=360$ GPa) \cite{Kazutaka2011A}, \cite{Szczesniak2013F}, and  ${\rm SiH_{4}(H_{2})_{2}}$ ($T_{C}\sim 107$ K for $p=250$ GPa) \cite{YLiPNAS}.
 
Historically, the physical properties of the ${\rm H_{2}S}$ compound have been studied for many years.
On the molecular level ${\rm H_{2}S}$ is formally the analogue of ${\rm H_{2}O}$. However, in the solid phase its properties are significantly different due to the fact that hydrogen sulfide is very weak hydrogen-bonded \cite{Ikeda2001A}.
It should be noted that ${\rm H_{2}S}$ has the complicated pressure-temperature phase diagram. In the area: $p\in\left<0,50\right>$ GPa and $T\in\left<0,300\right>$ K, as many as seven crystal structures are distinct \cite{Cockcroft1990A}, \cite{Shimizu1991A}, \cite{Shimizu1992A}, \cite{Endo1994A}, \cite{Endo1996A}, \cite{Fujihisa1998A}. On the other hand, the recently published theoretical results call into question the original findings on the number and the type of the existing crystal structures \cite{Li2014A}.
It is also worth noting that to the present date there is no final consensus on the pressure at which the molecular dissociation ${\rm H_{2}S}$ occurs at the room temperature. In literature one can find the two characteristic values of the pressure: $27$ GPa and $47$ GPa \cite{Sakashita1997A}, \cite{Fujihisa2004A}, \cite{Shimizu1997A}. 

The present paper determines the thermodynamic parameters of the superconducting state in the ${\rm H_{2}S}$ compound in the range of the pressures from $130$ GPa to $180$ GPa. Then, the study generalizes the results on the entire family of the compounds of the ${\rm H_{n}S}$-type (also additionally hydrogenated).
The numerical calculations were performed in the framework of the Eliashberg formalism due to the significant strong-coupling and retardation effects.


The Eliashberg equations for the order parameter function $\phi\left(\omega\right)$ and for the wave function renormalization factor $Z\left(\omega\right)$ take the form \cite{Marsiglio1988A}:
\begin{eqnarray}
\label{r1}
\phi\left(\omega\right)&=&
\frac{\pi}{\beta}\sum_{m=-M}^{M}\frac{\left[\lambda\left(\omega-i\omega_{m}\right)-\mu^{\star}\left(\omega_{m}\right)\right]}
{\sqrt{\omega_m^2Z^{2}_{m}+\phi^{2}_{m}}}\phi_{m}\\ \nonumber
                              &+& i\pi\int_{0}^{+\infty}d\omega^{'}\alpha^{2}F\left(\omega^{'}\right)
                                  \Big[\left[N\left(\omega^{'}\right)+f\left(\omega^{'}-\omega\right)\right]\\ \nonumber
                              &\times&K\left(\omega,-\omega^{'}\right)\phi\left(\omega-\omega^{'}\right)\Big]\\ \nonumber
                              &+& i\pi\int_{0}^{+\infty}d\omega^{'}\alpha^{2}F\left(\omega^{'}\right)
                                  \Big[\left[N\left(\omega^{'}\right)+f\left(\omega^{'}+\omega\right)\right]\\ \nonumber
                              &\times&K\left(\omega,\omega^{'}\right)\phi\left(\omega+\omega^{'}\right)\Big],
\end{eqnarray}
and
\begin{eqnarray}
\label{r2}
Z\left(\omega\right)&=&
                                  1+\frac{i\pi}{\omega\beta}\sum_{m=-M}^{M}
                                  \frac{\lambda\left(\omega-i\omega_{m}\right)\omega_{m}}{\sqrt{\omega_m^2Z^{2}_{m}+\phi^{2}_{m}}}Z_{m}\\ \nonumber
                              &+&\frac{i\pi}{\omega}\int_{0}^{+\infty}d\omega^{'}\alpha^{2}F\left(\omega^{'}\right)
                                  \Big[\left[N\left(\omega^{'}\right)+f\left(\omega^{'}-\omega\right)\right]\\ \nonumber
                              &\times&K\left(\omega,-\omega^{'}\right)\left(\omega-\omega^{'}\right)Z\left(\omega-\omega^{'}\right)\Big]\\ \nonumber
                              &+&\frac{i\pi}{\omega}\int_{0}^{+\infty}d\omega^{'}\alpha^{2}F\left(\omega^{'}\right)
                                  \Big[\left[N\left(\omega^{'}\right)+f\left(\omega^{'}+\omega\right)\right]\\ \nonumber
                              &\times&K\left(\omega,\omega^{'}\right)\left(\omega+\omega^{'}\right)Z\left(\omega+\omega^{'}\right)\Big], 
\end{eqnarray}
where:
\begin{equation}
\label{r3}
K\left(\omega,\omega^{'}\right)\equiv
\frac{1}{\sqrt{\left(\omega+\omega^{'}\right)^{2}Z^{2}\left(\omega+\omega^{'}\right)-\phi^{2}\left(\omega+\omega^{'}\right)}}.
\end{equation}
The order parameter is defined as: $\Delta\left(\omega\right)\equiv \phi\left(\omega\right)/Z\left(\omega\right)$. 

The imaginary axis functions ($\phi_{n}\equiv\phi\left(i\omega_{n}\right)$ and $Z_{n}\equiv Z\left(i\omega_{n}\right)$) should be calculated from \cite{Eliashberg1960A}: 
\begin{equation}
\label{r4}
\phi_{n}=\frac{\pi}{\beta}\sum_{m=-M}^{M}
\frac{\lambda\left(i\omega_{n}-i\omega_{m}\right)-\mu^{\star}\left(\omega_{m}\right)}
{\sqrt{\omega_m^2Z^{2}_{m}+\phi^{2}_{m}}}\phi_{m},
\end{equation}
and
\begin{equation}
\label{r5}
Z_{n}=1+\frac{1}{\omega_{n}}\frac{\pi}{\beta}\sum_{m=-M}^{M}
\frac{\lambda\left(i\omega_{n}-i\omega_{m}\right)}{\sqrt{\omega_m^2Z^{2}_{m}+\phi^{2}_{m}}}
\omega_{m}Z_{m},
\end{equation}
where: $\omega_{n}\equiv\left(\pi/\beta\right)\left(2n-1\right)$ is the Matsubara frequency and $\beta\equiv\left(k_{B}T\right)^{-1}$. The symbol $k_{B}$ represents the Boltzmann constant.
The pairing kernel is given by:
\begin{equation}
\label{r6}
\lambda\left(z\right)\equiv 2\int_0^{\Omega_{\rm{max}}}d\Omega\frac{\Omega}{\Omega ^2-z^{2}}\alpha^{2}F\left(\Omega\right).
\end{equation}
The Eliashberg function ($\alpha^{2}F\left(\Omega\right)$) models the structure of the electron-phonon interaction and $\Omega_{\rm{max}}$ is the maximum phonon frequency. For the superconductor ${\rm H_{2}S}$, in the range of the pressure from $130$ GPa to $180$ GPa, the Eliashberg functions were calculated in the paper \cite{Li2014A}. The maximum phonon frequency is in the order of $220$ meV.
The function $\mu^{\star}\left(\omega_{n}\right)\equiv\mu^{\star}\theta\left(\omega_{c}-|\omega_{n}|\right)$ describes the depairing Coulomb interaction, 
where $\mu^{\star}$ represents the Coulomb pseudopotential \cite{Morel1962A}. The symbol $\theta$ is the Heaviside unit function and $\omega_{c}$ denotes the cut-off energy ($\omega_{c}=3\Omega_{\rm{max}}$). 
The Bose function and the Fermi function is given by the symbol $N\left(\omega\right)$ and $f\left(\omega\right)$, respectively. 

The Eliashberg equations have been solved for $M=1100$. The functions $\phi\left(\omega\right)$ and $Z\left(\omega\right)$ are stable for 
$T\geq T_{0}\equiv 5$ K. Note that the above was based on the numerical methods used in the publications: \cite{Szczesniak2013H}, \cite{Szczesniak2014B}, \cite{Szczesniak2014C}, \cite{Szczesniak2014E}. 

\begin{figure}
\includegraphics*[width=\columnwidth]{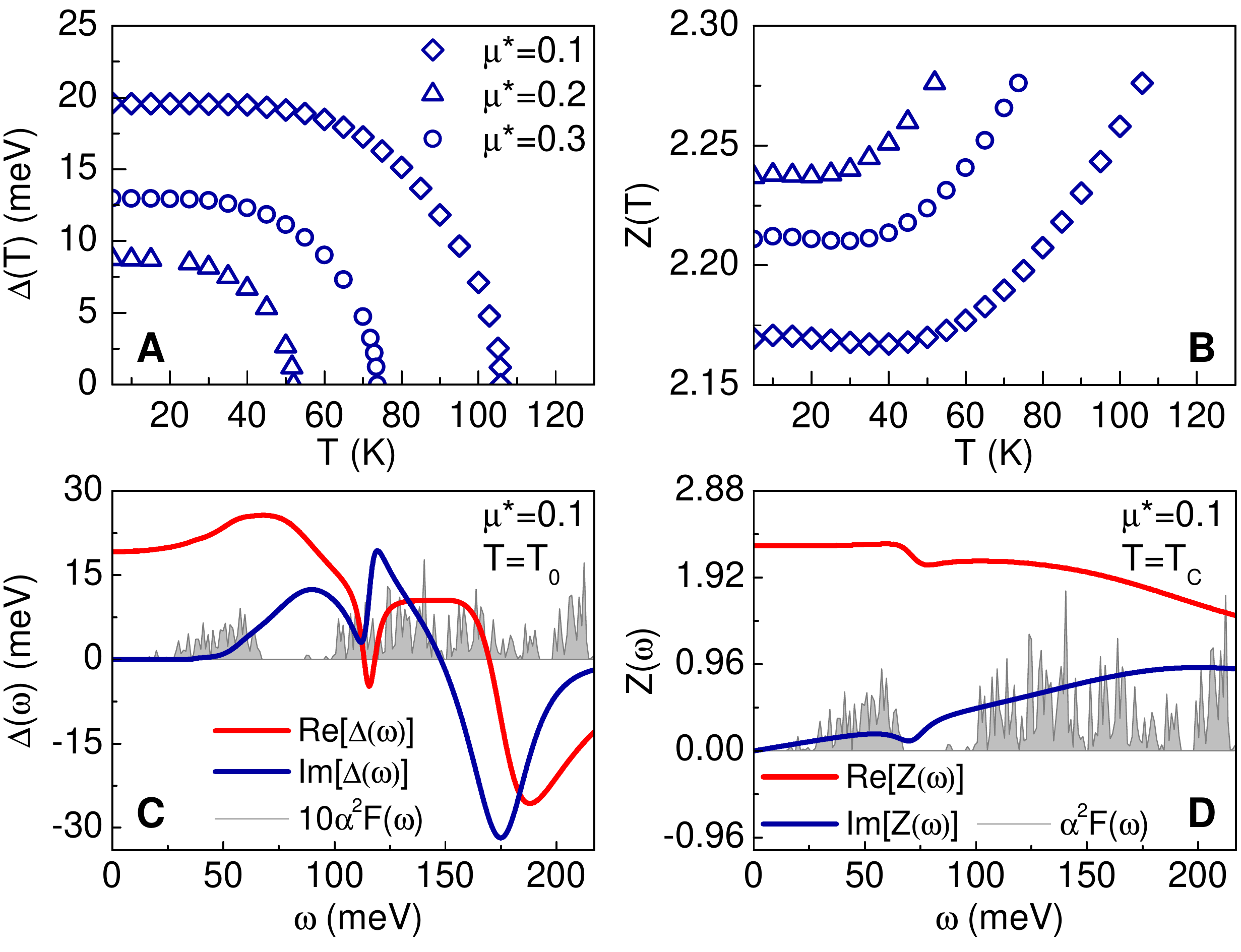}
\caption{The dependence of (A) the order parameter and (B) the wave function renormalization factor on the temperature for the selected values of the Coulomb pseudopotential. The figures (C)-(D) present the overt form of the functions $\Delta\left(\omega\right)$ and $Z\left(\omega\right)$ for the selected temperatures. The shaded area is the rescaled Eliashberg function.}
\label{f1} 
\end{figure}

\fig{f1} (A)-(B) shows the plots of the exemplary dependence of the order parameter and the wave function renormalization factor on the temperature ($p=160$ GPa). The wide range of the values of the Coulomb pseudopotential was taken into account. 
It turns out that the resulting curves can be reproduced with the very good approximation by using the functions: 
$\Delta\left(T,\mu^{\star}\right)=\Delta\left(\mu^{\star}\right)\sqrt{1-\left(\frac{T}{T_{C}}\right)^{\Gamma}}$ and
$Z\left(T,\mu^{\star}\right)=Z\left(\mu^{\star}\right)+\left[Z\left(T_{C}\right)-Z\left(\mu^{\star}\right)\right]\left(\frac{T}{T_{C}}\right)^{\Gamma}$, 
where: 
$\Delta\left(\mu^{\star}\right)=118.79\left(\mu^{\star}\right)^2-101.78\mu^{\star}+28.60$,  
$Z\left(\mu^{\star}\right)=-0.792\left(\mu^{\star}\right)^2+0.654\mu^{\star}+2.112$, 
$Z\left(T_{C}\right)\simeq 2.28=1+\lambda$, and $\Gamma=3.6$ (the electron-phonon coupling constant ($\lambda$) was defined in \tab{t1}). 

In the first case, it is clear that the high values of the order parameter correspond to a high values of the critical temperature (\fig{f1} (A)). It should be noted that the full shape of the function $\Delta\left(T,\mu^{\star}\right)$ cannot be properly identified within the framework of the classical BCS theory, as: $\left[\Gamma\right]_{\rm BCS}=3$ \cite{Eschrig2001A}.

On the other hand, the wave function renormalization factor determines the value of the electron effective mass: $m^{\star}_{e}=Z\left(T\right)m_{e}$, where $m_{e}$ denotes the electron band mass. 
The results plotted in \fig{f1} (B) prove that the effective mass of the electrons is large in the superconducting state. From the physical point of view the obtained result comes from the existence of the significant strong-coupling effects in the ${\rm H_{2}S}$ compound, which are characterized by the electron-phonon coupling constant. Of course, these effects cannot be ignored in the quantitative analysis. Let us notice that the BCS model predicts: $m^{\star}_{e}=m_{e}$.
      
The curves in \fig{f1} (A) and (B) were obtained on the basis of the following expressions: 
$\Delta\left(T\right)={\rm Re}\left[\Delta\left(\omega=\Delta\left(T\right),T\right)\right]$ and 
$Z\left(T\right)={\rm Re}\left[Z\left(\omega=0,T\right)\right]$, while the example solutions of the Eliashberg equations are presented in \fig{f1} (C) and (D). 
It can be seen that the order parameter has the zero imaginary part at the low frequencies, due to the absence of the damping effects \cite{Varelogiannis1997A}. At the higher frequencies both $\Delta\left(\omega\right)$ and $Z\left(\omega\right)$ are characterized by the very complicated shape clearly correlated with the shape of the Eliashberg function. 

\begin{figure}
\includegraphics*[width=\columnwidth]{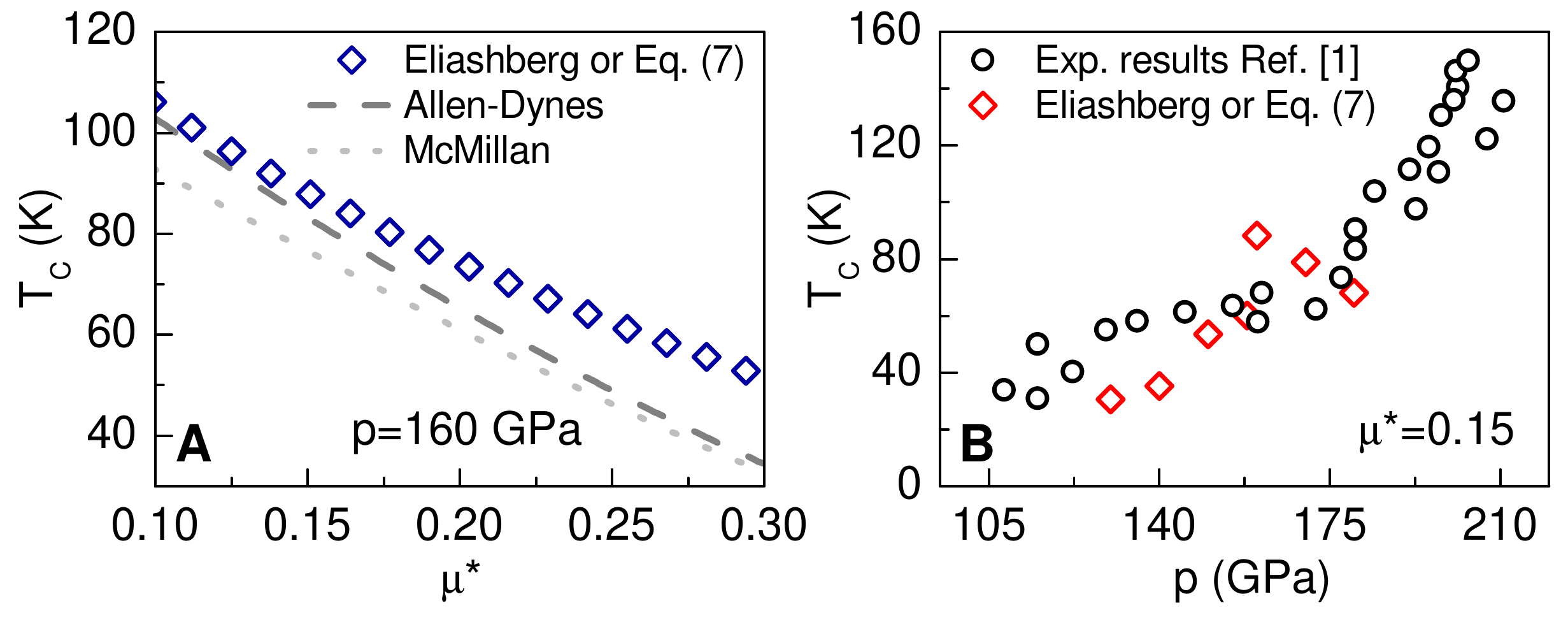}
\caption{The critical temperature as a function of (A) the Coulomb pseudopotential for $p=160$ GPa and (B) the pressure for $\mu^{\star}=0.15$. The experimental results are taken from \cite{Drozdov2014A}.}
\label{f2} 
\end{figure}

\fig{f2} (A) presents the influence of the Coulomb pseudopotential on the value of the critical temperature in the ${\rm H_{2}S}$ compound. It can be seen that in the case of the weak electron depairing correlations ($\mu^{\star}\sim 0.1$), $T_{C}$ can reach the high value of the order of $100$ K.
  
The numerical results obtained with the help of the Eliashberg equations can be reproduced with the very good accuracy using the modified Allen-Dynes formula:
\begin{equation}
\label{r7}
k_{B}T_{C}=f_{1}f_{2}\frac{\omega_{{\rm ln}}}{1.27}\exp\left[\frac{-1.14\left(1+\lambda\right)}{\lambda-(1 + 0.163 \lambda)\mu^{\star}}\right],
\end{equation}
whereas the symbols appearing in \eq{r7} were defined in \tab{t1}. In contrast, the numerical parameters were selected using the method of the least squares on the basis of $300$ numerical values of the function $T_{C}\left(\mu^{\star}\right)$.

\begin{figure}
\includegraphics*[width=0.8\columnwidth]{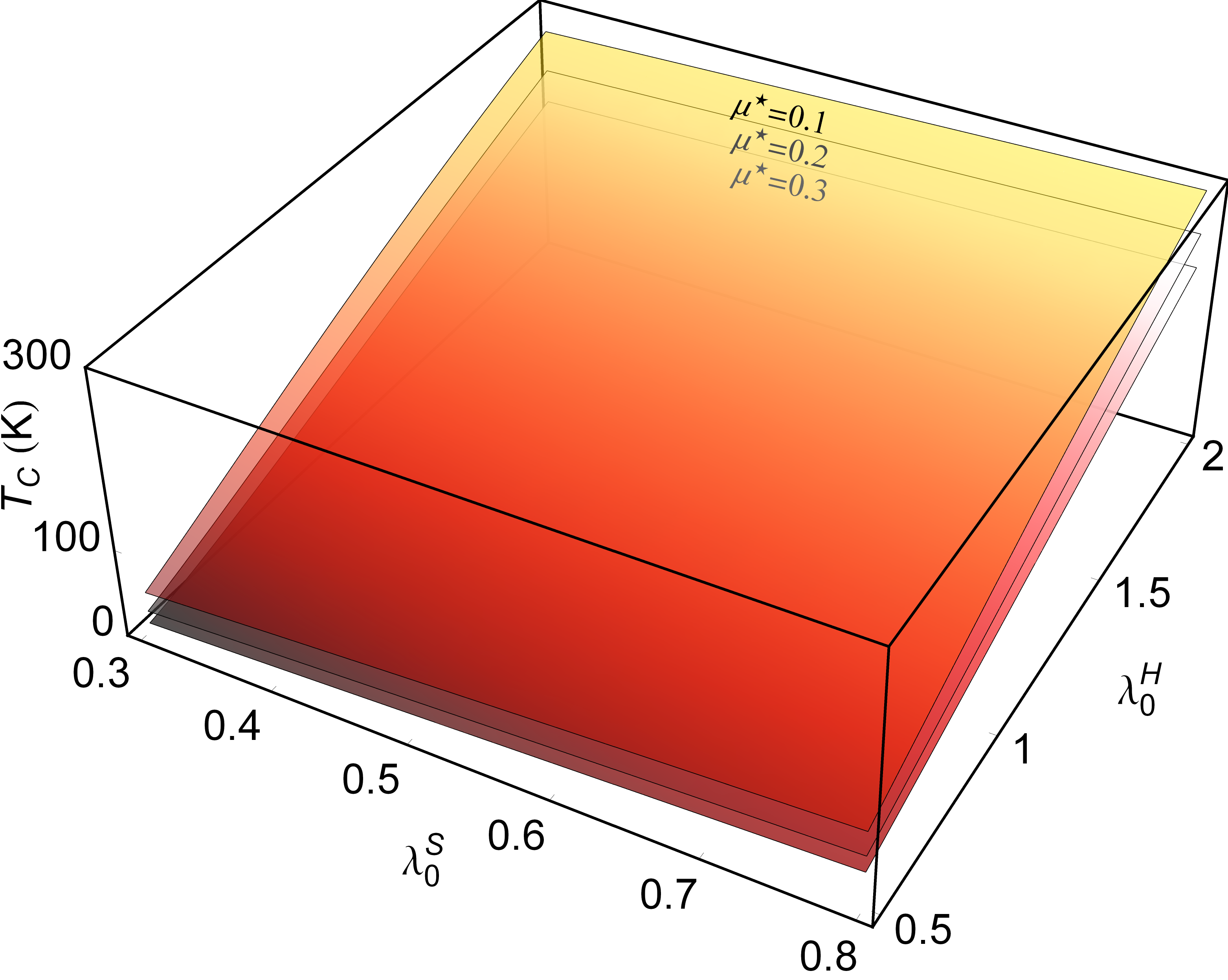}
\caption{ The critical temperature as a function of the coupling constants $\lambda^{\rm S}_{0}$ and $\lambda^{\rm H}_{0}$ for the superconductors 
of the ${\rm H_{n}S}$-type.}
\label{f3} 
\end{figure}
%

%
\begin{table}
\caption{\label{t1}
The quantities $\lambda$, $\omega_{{\rm ln}}$, and $\omega_{2}$ represent the electron-phonon coupling constant, the logarithmic phonon frequency, and the second moment of the normalized weight function. The parameters $f_{1}$ and $f_{2}$ are the strong-coupling correction function and the shape correction function, respectively \cite{Allen1975A}.}
\begin{ruledtabular}
\begin{tabular}{ll}
Quantity & Value ($p=160$ GPa)\\
\hline
 & \\
${\lambda\equiv 2\int^{+\infty}_0 d\Omega \frac{\alpha^2\left(\Omega\right)F\left(\Omega\right)}{\Omega}}$& 1.28\\
$\omega_{{\rm ln}}\equiv \exp\left[\frac{2}{\lambda}
\int^{+\infty}_{0}d\Omega\frac{\alpha^{2}F\left(\Omega\right)}
{\Omega}\ln\left(\Omega\right)\right]$ & 82.70 meV\\
 & \\ 
$\sqrt{\omega_{2}}\equiv 
\left[\frac{2}{\lambda}
\int^{+\infty}_{0}d\Omega\alpha^{2}F\left(\Omega\right)\Omega\right]^{1/2}$ & 115.05 meV\\
& \\
$f_{1}\equiv\left[1+\left(\frac{\lambda}{\Lambda_{1}}\right)^{\frac{3}{2}}\right]^{\frac{1}{3}}$ & - \\
 & \\
$f_{2}\equiv 1+\frac
{\left(\frac{\sqrt{\omega_{2}}}{\omega_{\rm{ln}}}-1\right)\lambda^{2}}
{\lambda^{2}+\Lambda^{2}_{2}}$ & - \\
 & \\
$\Lambda_{1}\equiv 2.4-0.14\mu^{\star}$ & - \\
 & \\
$\Lambda_{2}\equiv\left(0.1+9\mu^{\star}\right)\left(\sqrt{\omega_{2}}/\omega_{\ln}\right)$ & - \\
 & \\
\end{tabular}
\end{ruledtabular}
\end{table}
%

Additionally, \fig{f2} (A) shows the plot of the values of the critical temperature obtained with the help of the classical Allen-Dynes and McMillan formulas \cite{Allen1975A}, \cite{McMillan1968A}. It was found that the classical formulas significantly understate $T_{C}$ for the higher values of the Coulomb pseudopotential.

\fig{f2} (B) presents the experimental dependence of the critical temperature on the pressure for the compound ${\rm H_{2}S}$ \cite{Drozdov2014A}. Note that the obtained results can be reproduced using the Eliashberg equations or \eq{r7} adopting: $\mu^{\star}=0.15$. In particular, for the pressure values from $130$ GPa to $180$ GPa, it was obtained: $T_{C}\in\left<31,88\right>$ K. 

Let us notice that \eq{r7} allows to discuss possible to achieve values of the critical temperature for the whole family of 
the compounds of the ${\rm H_{n}S}$-type.
In the first step it should be noted that contributions to the Eliashberg function derived from sulfur and hydrogen (both for ${\rm H_{2}S}$ and $\rm(H_{2}S)_{2}H_{2}$) are very clearly separated \cite{Duan2014A}, \cite{Li2014A}. In particular, in the frequency range from $0$ to $\sim 70$ meV the crucial is the electron-phonon interaction derived from sulfur, while above $\sim 100$ meV significant is the contribution derived from hydrogen. 
Based on the above fact, the model Eliashberg function was factorized as follows:
\begin{eqnarray}
\label{r8}
\alpha^{2}F\left(\Omega\right)&=&\lambda^{\rm S}_{0}\left(\frac{\Omega}{\Omega^{\rm S}_{\rm max}}\right)^{2}\theta\left(\Omega^{\rm S}_{\rm max}-\Omega\right)
\\ \nonumber
&+&
\lambda^{\rm H}_{0}\left(\frac{\Omega}{\Omega^{\rm H}_{\rm max}}\right)^{2}\theta\left(\Omega^{\rm H}_{\rm max}-\Omega\right),
\end{eqnarray}
where $\lambda^{\rm S}_{0}$ and $\lambda^{\rm H}_{0}$ are the contributions to the electron-phonon coupling constant derived respectively from sulfur and hydrogen. On the other hand, the symbols $\Omega^{\rm S}_{\rm max}$ and $\Omega^{\rm H}_{\rm max}$ represent the maximum phonon frequencies. 

Using the equations included in \tab{t1}, it can be shown that:
\begin{equation}
\label{r9}
\lambda=\lambda^{\rm S}_{0}+\lambda^{\rm H}_{0},
\end{equation}
\begin{eqnarray}
\label{r10}
\omega_{\rm ln}&=&{\rm exp}\left[\frac{\lambda^{\rm S}_{0}}{\lambda^{\rm S}_{0}+\lambda^{\rm H}_{0}}\left(\ln\left(\Omega^{\rm S}_{\rm max}\right)-\frac{1}{2}\right)\right]\\ \nonumber
&\times&
{\rm exp}\left[\frac{\lambda^{\rm H}_{0}}{\lambda^{\rm S}_{0}+\lambda^{\rm H}_{0}}\left(\ln\left(\Omega^{\rm H}_{\rm max}\right)-\frac{1}{2}\right)\right],
\end{eqnarray}
and
\begin{equation}
\label{r11}
\omega_{2}=\frac{\lambda^{\rm S}_{0}}{\lambda^{\rm S}_{0}+\lambda^{\rm H}_{0}}\frac{\left(\Omega^{\rm S}_{\rm max}\right)^{2}}{2}
+
\frac{\lambda^{\rm H}_{0}}{\lambda^{\rm S}_{0}+\lambda^{\rm H}_{0}}\frac{\left(\Omega^{\rm H}_{\rm max}\right)^{2}}{2}.
\end{equation}
\begin{figure}
\includegraphics*[width=0.95\columnwidth]{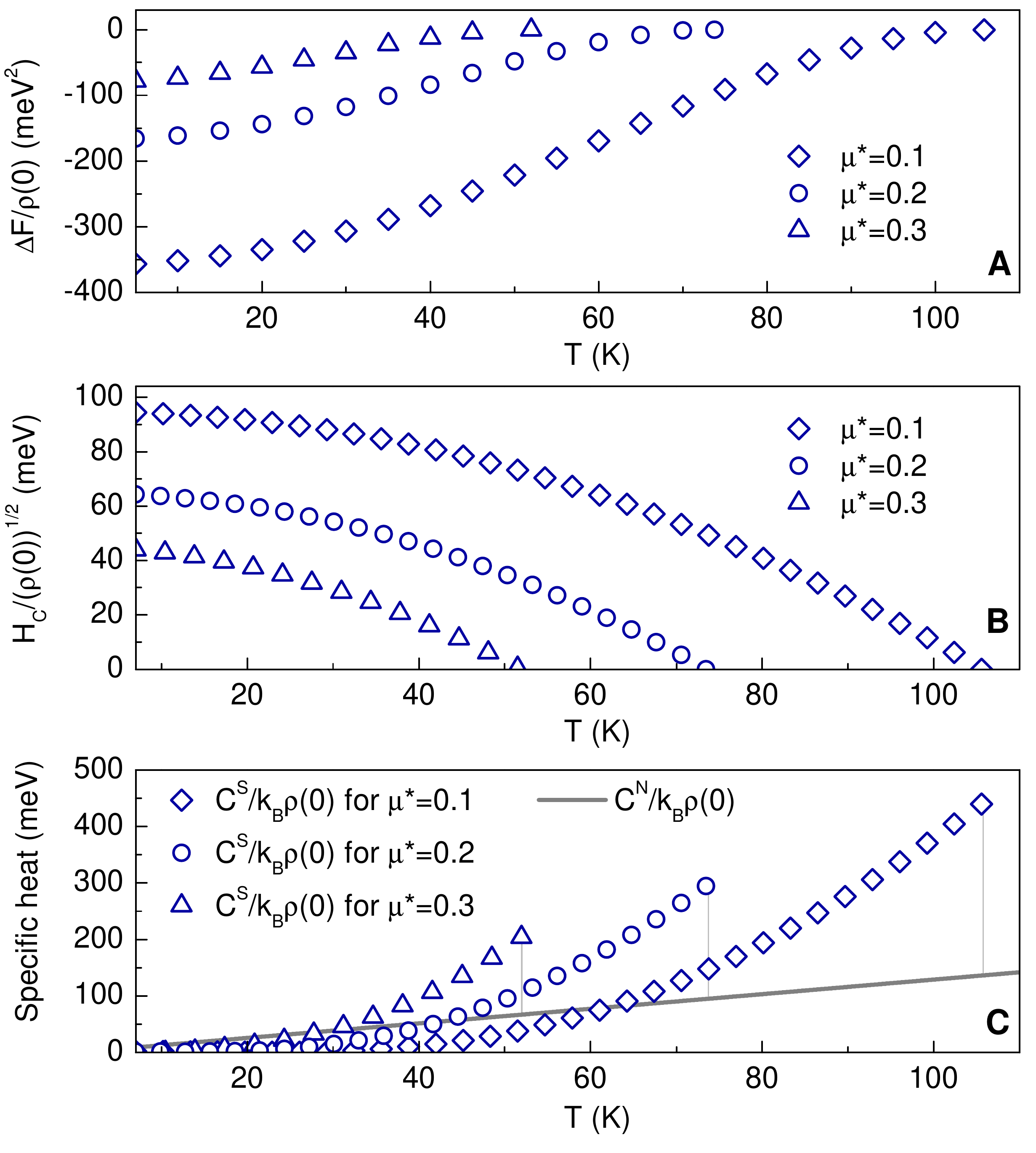}
\caption{(A) The free energy difference between the superconducting and normal state, (B) the thermodynamic critical field, and (C) the specific heat of the superconducting state $C^{S}$ and the normal state $C^{N}$ as a function of the temperature for the selected values of the Coulomb pseudopotential. The vertical lines indicate the position of the characteristic specific heat jump at $T_C$.
}
\label{f4} 
\end{figure}

\fig{f3} presents the dependence of the critical temperature on $\lambda^{\rm S}_{0}$ and $\lambda^{\rm H}_{0}$. It was adopted: 
$\lambda^{\rm S}_{0}\in\left<0.3,0.8\right>$, $\lambda^{\rm H}_{0}\in\left<0.5,2\right>$, $\Omega^{\rm S}_{\rm max}=70$ meV, $\Omega^{\rm H}_{\rm max}=220$ meV \cite{Szczesniak2012K}, \cite{Szczesniak2012L}, and the selected values of $\mu^{\star}$. 

The obtained results show that for the low values of the Coulomb pseudopotential, the maximum critical temperature can be equal even to $\sim 290$ K. 
From the physical point of view, the result shows the possibility of induction of the superconducting state with the critical temperature comparable to the room temperature in the compounds of the ${\rm H_{n}S}$-type. 

The thermodynamic critical field and the specific heat of the superconducting state can be calculated in the Eliashberg formalism using the formula for the free energy difference between the superconducting state ($S$) and the normal state ($N$): 
\begin{eqnarray}
\label{r12}
\frac{\Delta F}{\rho\left(0\right)}&=&-\frac{2\pi}{\beta}\sum_{n=1}^{M}
\left(\sqrt{\omega^{2}_{n}+\Delta^{2}_{n}}- \left|\omega_{n}\right|\right)\\ \nonumber
&\times&(Z^{S}_{n}-Z^{N}_{n}\frac{\left|\omega_{n}\right|}
{\sqrt{\omega^{2}_{n}+\Delta^{2}_{n}}}),
\end{eqnarray}  
where $\rho\left(0\right)$ denotes the value of the electron density of states on the Fermi level.
 
The formula for the thermodynamic critical field has the form: $H_{C}=\sqrt{-8\pi\Delta F}$. On the other hand, the specific heat of the superconducting 
state can be expressed using the formula: $C^{S}=\Delta C+C^{N}$, where $\Delta C=-T d^{2}\Delta F/dT^{2}$ and $C^{N}\left(T\right)=\gamma T$. The symbol $\gamma$ denotes the Sommerfeld constant: $\gamma\equiv({2}/{3})\pi^{2}k^{2}_{B}\rho\left(0\right)\left(1+\lambda\right)$.

\begin{figure}
\includegraphics*[width=\columnwidth]{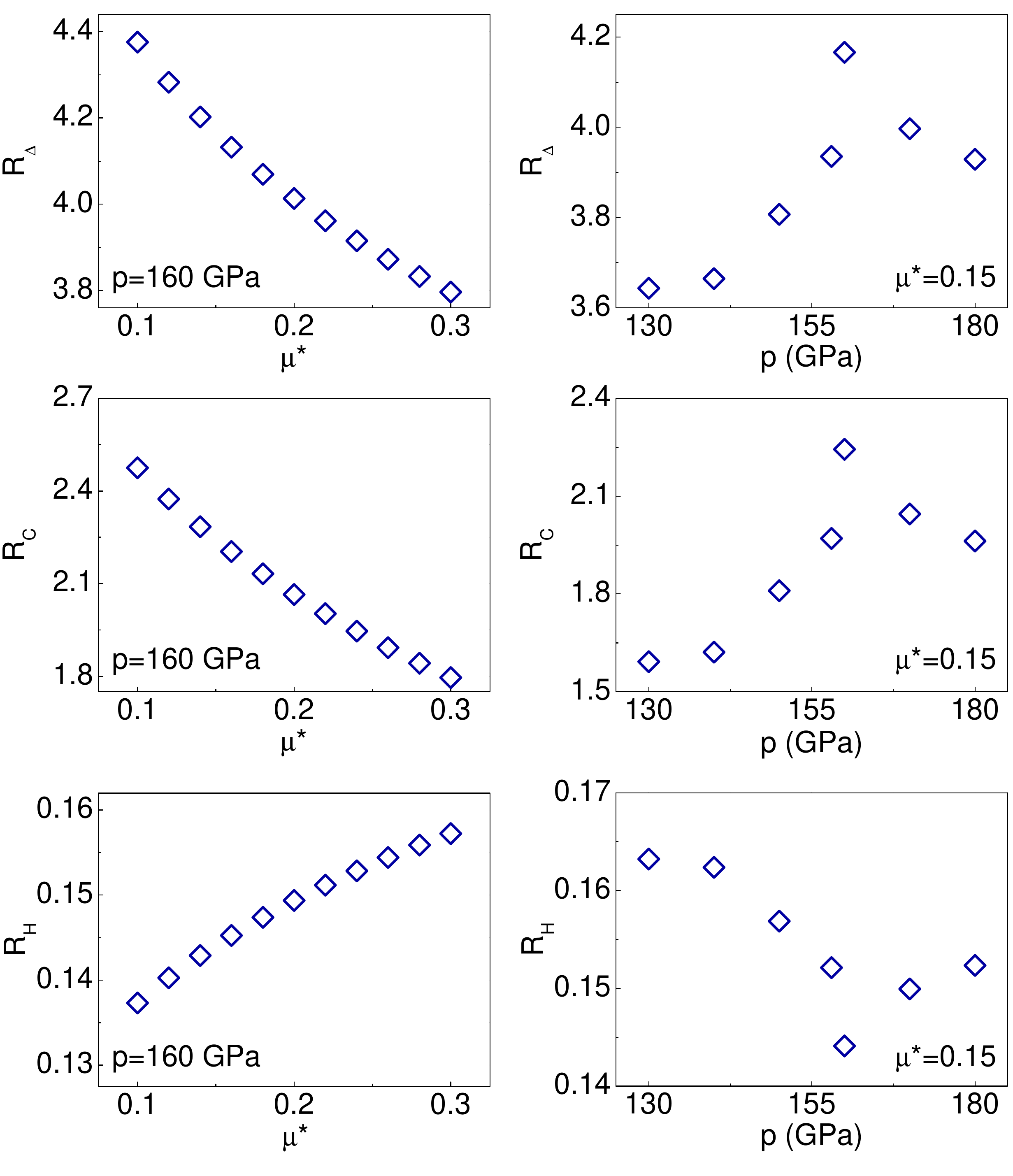}
\caption{The dimensionless thermodynamic ratios as a function of the Coulomb pseudopotential for $p=160$ GPa (left panel) and as a function of the pressure for $\mu^{\star}=0.15$ (right panel).}
\label{f5} 
\end{figure}
\begin{figure}
\includegraphics*[width=0.75\columnwidth]{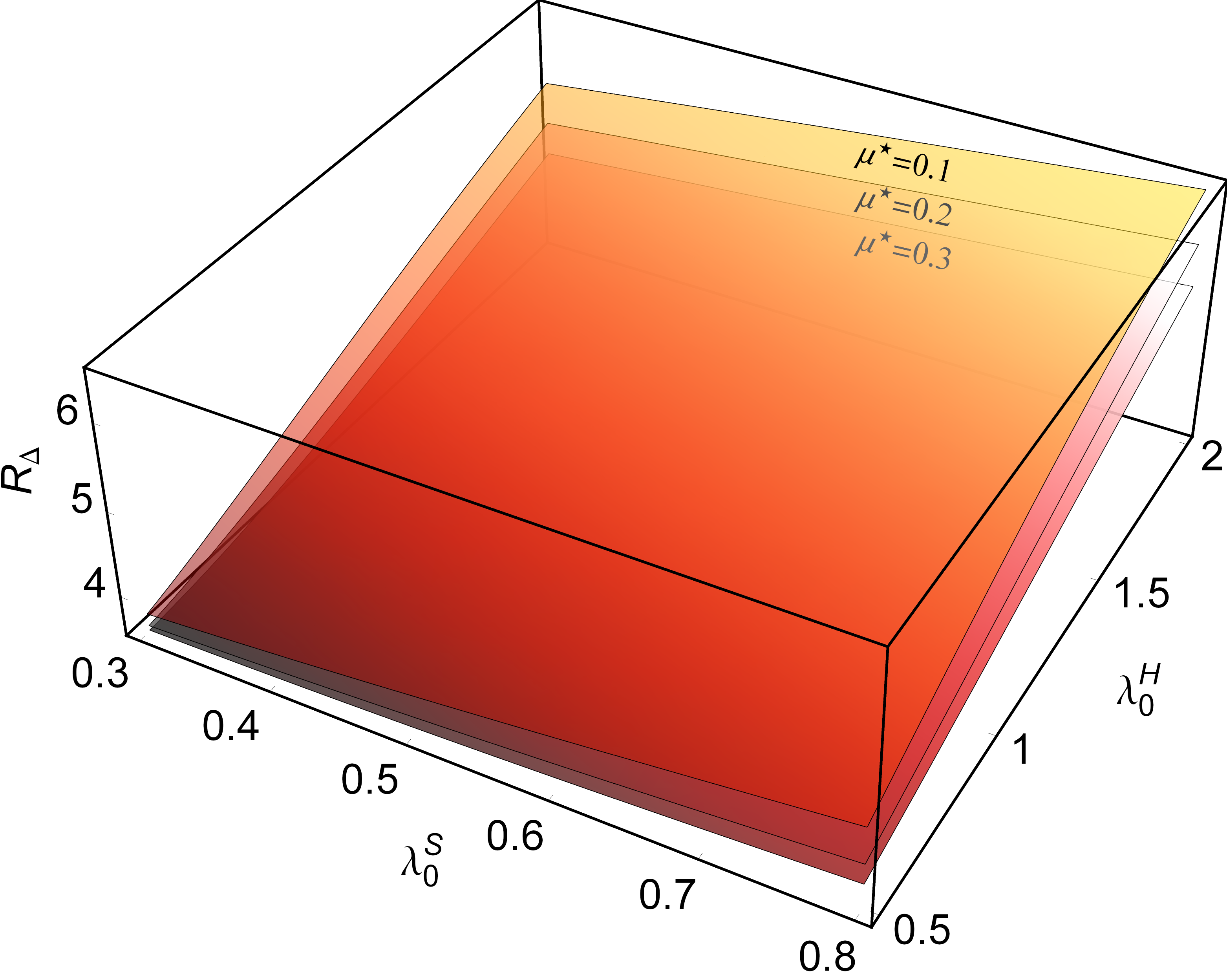}
\includegraphics*[width=0.75\columnwidth]{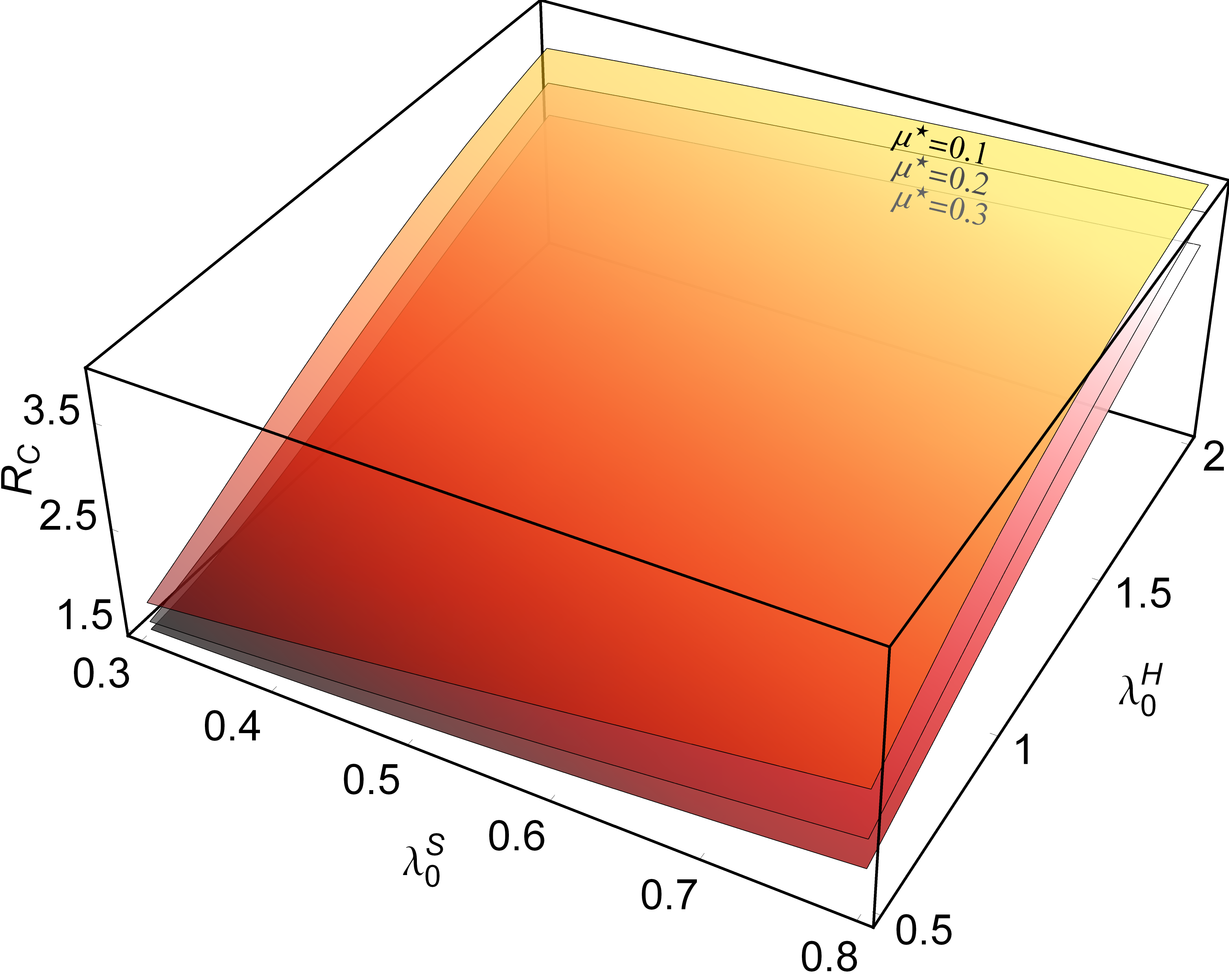}
\includegraphics*[width=0.75\columnwidth]{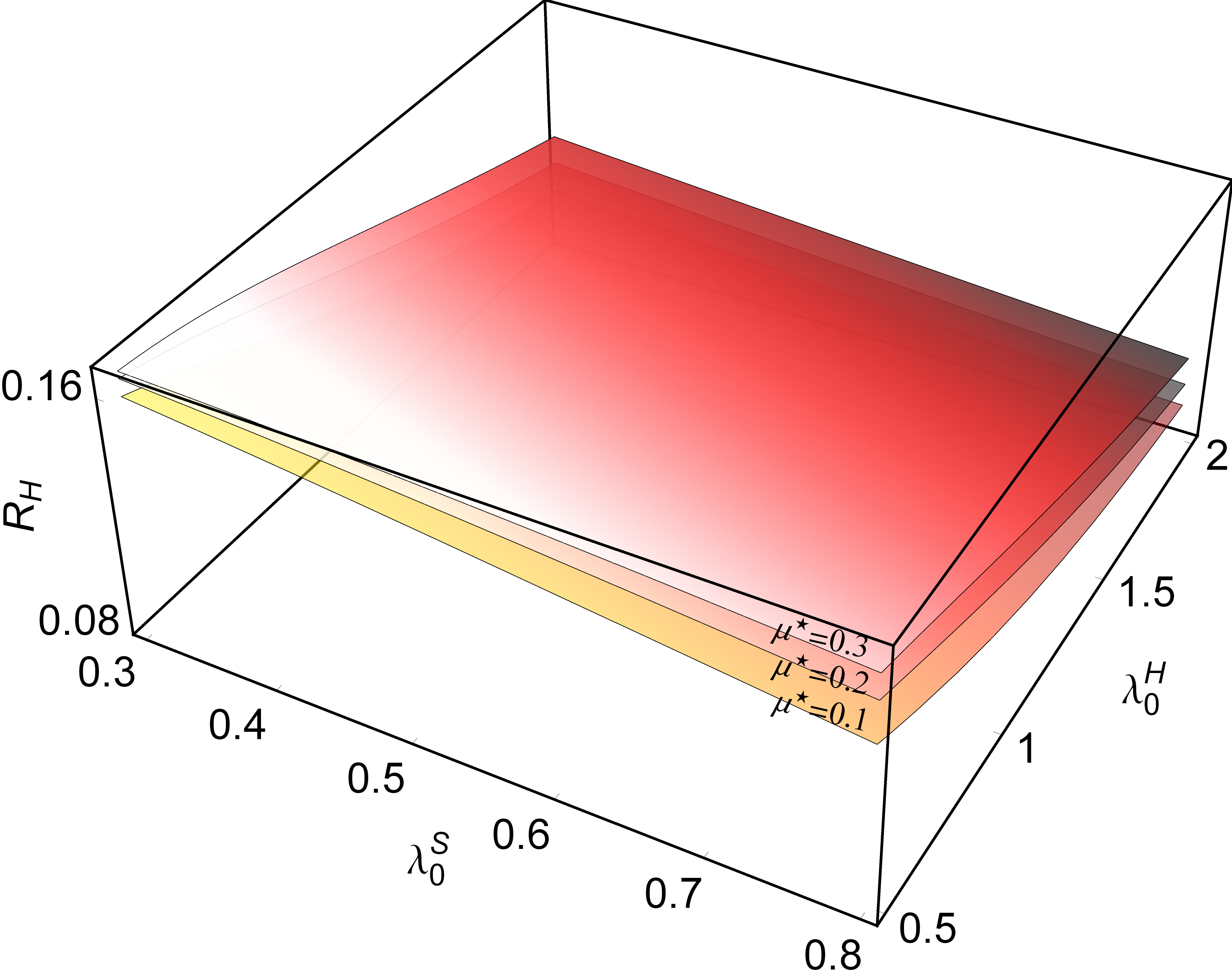}
\caption{The dimensionless thermodynamic ratios as a function of the coupling constants $\lambda^{\rm S}_{0}$ and $\lambda^{\rm H}_{0}$ 
for the ${\rm H_{n}S}$-type superconductors.}
\label{f6} 
\end{figure}

The examples of the results obtained for $p=160$ GPa are plotted in \fig{f4}. It can be seen that below the critical temperature, the difference in the free energy takes the negative values, which from the physical point of view shows the thermodynamic stability of the superconducting phase (\fig{f4} (A)). 
It should also be noted that the values of $\Delta F$ strongly depend on the Coulomb pseudopotential.
The strong dependence of the free energy difference on the Coulomb pseudopotential transfers directly to the thermodynamic critical field and the specific heat of the superconducting state (see \fig{f4} (B) and (C)). In particular, it was obtained: 
$\left[H_{C}\left(0\right)\right]_{\mu^{\star}=0.1}/\left[H_{C}\left(0\right)\right]_{\mu^{\star}=0.3}=2.16$ and 
$\left[\Delta C\left(T_{C}\right)\right]_{\mu^{\star}=0.1}/\left[\Delta C\left(T_{C}\right)\right]_{\mu^{\star}=0.3}=2.21$, 
while $H_{C}\left(0\right)\equiv H_{C}\left(T_{0}\right)$. 

The designated thermodynamic functions allow to calculate the dimensionless ratios: $R_{\Delta}\equiv 2\Delta(0)/k_{B}T_{C}$, 
$R_{C}\equiv{\Delta C\left(T_{C}\right)}/{C^{N}\left(T_{C}\right)}$, and $R_{H}\equiv{T_{C}C^{N}\left(T_{C}\right)}/{H_{C}^{2}\left(0\right)}$, where $\Delta\left(0\right)\equiv \Delta\left(T_{0}\right)$.
It is worth noting that, in the BCS theory, the quantities: $R_{\Delta}$, $R_{C}$, and $R_{H}$ adopt the universal values, which are equal respectively to: $3.53$, $1.43$, and $0.168$ \cite{Bardeen1957A}, \cite{Bardeen1957B}. In the case of ${\rm H_{2}S}$ ($p=160$ GPa), the obtained results are presented 
in \fig{f5}. It is easy to note that the values $R_{\Delta}$ - $R_{H}$ differ significantly from the predictions of the classical BCS theory, with the largest derogations observed in the case of the weak electron depairing correlations.

\fig{f5} shows the plots of the dimensionless ratios as a function of the pressure ($p\in\left<130, 180\right>$ GPa). The numerical results show that the thermodynamic parameters $R_{\Delta}$ - $R_{H}$ cannot be correctly estimated in the framework of the BCS theory for the wide range of pressure: $R_{\Delta}\in\left<3.64,4.16\right>$, $R_{C}\in\left<1.59,2.24\right>$, and $R_{H}\in\left<0.144, 0.163\right>$. 
From the physical standpoint, the results presented above arise from the existence of the strong-coupling and retardation effects in ${\rm H_{2}S}$. Note that in the simplest case, it can be characterized by the ratio: $k_{B}T_{C}/\omega_{\rm ln}$ \cite{Carbotte1990A}. Thus, the results included in \fig{f5} can be reproduced with the help of the following formulas:
\begin{equation}
\label{r13}
\frac{R_{\Delta}}{\left[R_{\Delta}\right]_{{\rm BCS}}}
=1+9\left(\frac{k_{B}T_{C}}{\omega_{{\rm ln}}}\right)^{2}\ln\left(\frac{\omega_{{\rm \ln}}}{k_{B}T_{C}}\right),
\end{equation}
\begin{equation}
\label{r14}
\frac{R_{C}}{\left[R_{C}\right]_{{\rm BCS}}}
=1+40\left(\frac{k_{B}T_{C}}{\omega_{{\rm ln}}}\right)^{2}\ln\left(\frac{\omega_{{\rm \ln}}}{2k_{B}T_{C}}\right),
\end{equation}
and
\begin{equation}
\label{r15}
\frac{R_{H}}{\left[R_{H}\right]_{{\rm BCS}}}
=1-10\left(\frac{k_{B}T_{C}}{\omega_{{\rm ln}}}\right)^{2}\ln\left(\frac{\omega_{{\rm \ln}}}{2k_{B}T_{C}}\right).
\end{equation}

It should be underlined that the number parameters appearing in \eq{r13} - \eq{r15} were selected on the basis of $300$ numerical values of $R_{\Delta}$ - $R_{H}$ depending on the Coulomb pseudopotential. 
\eq{r13} - \eq{r15} can be used to estimate $R_{\Delta}$ - $R_{H}$ for the whole family of the ${\rm H_{n}S}$-type compounds. The obtained results are plotted in \fig{f6}. It was found that the values, which maximally differ from the results of the BCS theory, are respectively: $6.53$, $3.99$, and $0.093$. 

%
%
\vspace*{0.25 cm}
In the presented paper, the thermodynamic parameters of the superconducting state inducing in the ${\rm H_{2}S}$ compound were determined 
($p\in\left<130, 180\right>$ GPa). The calculations were carried out in the framework of the Eliashberg formalism.
It has been shown that the theoretical analysis is able to reproduce the experimental dependence of the critical temperature on the pressure, assuming the relatively low value of the electron depairing correlations ($\mu^{\star}=0.15$).
 
The other thermodynamic functions such as: the order parameter, the thermodynamic critical field or the specific heat of the superconducting state deviate from the expectations of the classical BCS theory (as evidenced by the values of the parameters $R_{\Delta}$ - $R_{H}$). It turns out that this result is associated with the existence of the significant strong-coupling and retardation effects appearing in the ${\rm H_{2}S}$ compound. 
 
Generalizing the obtained results, we showed that the maximum value of the critical temperature in the family of the compounds of the ${\rm H_{n}S}$-type can be equal even to $\sim 290$ K, which from the physical point of view means the possibility of the existence of the superconducting phase at the room temperature. 
The values of the parameters $R_{\Delta}$ - $R_{H}$ for the family of ${\rm H_{n}S}$ differ also significantly from the expectations of the BCS theory.
%
\vspace*{-0.65 cm} 
\begin{acknowledgements}
Y. L. acknowledges funding from the National Natural Science Foundation of China under Grant Nos. 11204111, 11404148, 
the Natural Science Foundation of Jiangsu province under Grant No. BK20130223, and the PAPD of Jiangsu Higher Education Institutions. 
\end{acknowledgements}
%
%

\bibliography{bibliography}

\begin{thebibliography}{44}
\expandafter\ifx\csname natexlab\endcsname\relax\def\natexlab#1{#1}\fi
\expandafter\ifx\csname bibnamefont\endcsname\relax
  \def\bibnamefont#1{#1}\fi
\expandafter\ifx\csname bibfnamefont\endcsname\relax
  \def\bibfnamefont#1{#1}\fi
\expandafter\ifx\csname citenamefont\endcsname\relax
  \def\citenamefont#1{#1}\fi
\expandafter\ifx\csname url\endcsname\relax
  \def\url#1{\texttt{#1}}\fi
\expandafter\ifx\csname urlprefix\endcsname\relax\def\urlprefix{URL }\fi
\providecommand{\bibinfo}[2]{#2}
\providecommand{\eprint}[2][]{\url{#2}}

\bibitem[{\citenamefont{Drozdov et~al.}(2014)\citenamefont{Drozdov, Eremets,
  and Troyan}}]{Drozdov2014A}
\bibinfo{author}{\bibfnamefont{A.~P.} \bibnamefont{Drozdov}},
  \bibinfo{author}{\bibfnamefont{M.~I.} \bibnamefont{Eremets}},
  \bibnamefont{and} \bibinfo{author}{\bibfnamefont{I.~A.}
  \bibnamefont{Troyan}}, \bibinfo{journal}{arXiv:1412.0460}
  (\bibinfo{year}{2014}).

\bibitem[{\citenamefont{Hirsch and Marsiglio}(2015)}]{HirschH2S}
\bibinfo{author}{\bibfnamefont{J.~E.} \bibnamefont{Hirsch}} \bibnamefont{and}
  \bibinfo{author}{\bibfnamefont{F.}~\bibnamefont{Marsiglio}},
  \bibinfo{journal}{Physica C} \textbf{\bibinfo{volume}{511}},
  \bibinfo{pages}{45} (\bibinfo{year}{2015}).

\bibitem[{\citenamefont{Bernstein et~al.}(2015)\citenamefont{Bernstein,
  Hellberg, Johannes, Mazin, and Mehl}}]{BernsteinH2S}
\bibinfo{author}{\bibfnamefont{N.}~\bibnamefont{Bernstein}},
  \bibinfo{author}{\bibfnamefont{C.~S.} \bibnamefont{Hellberg}},
  \bibinfo{author}{\bibfnamefont{M.~D.} \bibnamefont{Johannes}},
  \bibinfo{author}{\bibfnamefont{I.~I.} \bibnamefont{Mazin}}, \bibnamefont{and}
  \bibinfo{author}{\bibfnamefont{M.~J.} \bibnamefont{Mehl}},
  \bibinfo{journal}{Phys. Rev. B} \textbf{\bibinfo{volume}{91}},
  \bibinfo{pages}{060511} (\bibinfo{year}{2015}).

\bibitem[{\citenamefont{Papaconstantopoulos
  et~al.}(2015)\citenamefont{Papaconstantopoulos, Klein, Mehl, and
  Pickett}}]{PapaconstantopoulosH2S}
\bibinfo{author}{\bibfnamefont{D.~A.} \bibnamefont{Papaconstantopoulos}},
  \bibinfo{author}{\bibfnamefont{B.~M.} \bibnamefont{Klein}},
  \bibinfo{author}{\bibfnamefont{M.~J.} \bibnamefont{Mehl}}, \bibnamefont{and}
  \bibinfo{author}{\bibfnamefont{W.~E.} \bibnamefont{Pickett}},
  \bibinfo{journal}{arXiv:1501.03950}  (\bibinfo{year}{2015}).

\bibitem[{\citenamefont{Flores-Livas et~al.}(2015)\citenamefont{Flores-Livas,
  Sanna, and Gross}}]{Flores-LivasH2S}
\bibinfo{author}{\bibfnamefont{J.~A.} \bibnamefont{Flores-Livas}},
  \bibinfo{author}{\bibfnamefont{A.}~\bibnamefont{Sanna}}, \bibnamefont{and}
  \bibinfo{author}{\bibfnamefont{E.~K.~U.} \bibnamefont{Gross}},
  \bibinfo{journal}{arXiv:1501.06336}  (\bibinfo{year}{2015}).

\bibitem[{\citenamefont{Akashi et~al.}(2015)\citenamefont{Akashi, Kawamura,
  Tsuneyuki, Nomura, and Arita}}]{AkashiH2S}
\bibinfo{author}{\bibfnamefont{R.}~\bibnamefont{Akashi}},
  \bibinfo{author}{\bibfnamefont{M.}~\bibnamefont{Kawamura}},
  \bibinfo{author}{\bibfnamefont{S.}~\bibnamefont{Tsuneyuki}},
  \bibinfo{author}{\bibfnamefont{Y.}~\bibnamefont{Nomura}}, \bibnamefont{and}
  \bibinfo{author}{\bibfnamefont{R.}~\bibnamefont{Arita}},
  \bibinfo{journal}{arXiv:1502.00936}  (\bibinfo{year}{2015}).

\bibitem[{\citenamefont{Zhang et~al.}(2015)\citenamefont{Zhang, Wang, Liu,
  Yang, Zhang, and Ma}}]{ZhangH2S}
\bibinfo{author}{\bibfnamefont{S.}~\bibnamefont{Zhang}},
  \bibinfo{author}{\bibfnamefont{Y.}~\bibnamefont{Wang}},
  \bibinfo{author}{\bibfnamefont{H.}~\bibnamefont{Liu}},
  \bibinfo{author}{\bibfnamefont{G.}~\bibnamefont{Yang}},
  \bibinfo{author}{\bibfnamefont{L.}~\bibnamefont{Zhang}}, \bibnamefont{and}
  \bibinfo{author}{\bibfnamefont{Y.}~\bibnamefont{Ma}},
  \bibinfo{journal}{arXiv:1502.02607}  (\bibinfo{year}{2015}).

\bibitem[{\citenamefont{Duan et~al.}(2014{\natexlab{a}})\citenamefont{Duan,
  Liu, Tian, Li, Huang, Zhao, Yu, Liu, Tian, and Cui}}]{Duan2014A}
\bibinfo{author}{\bibfnamefont{D.}~\bibnamefont{Duan}},
  \bibinfo{author}{\bibfnamefont{Y.}~\bibnamefont{Liu}},
  \bibinfo{author}{\bibfnamefont{F.}~\bibnamefont{Tian}},
  \bibinfo{author}{\bibfnamefont{D.}~\bibnamefont{Li}},
  \bibinfo{author}{\bibfnamefont{X.}~\bibnamefont{Huang}},
  \bibinfo{author}{\bibfnamefont{Z.}~\bibnamefont{Zhao}},
  \bibinfo{author}{\bibfnamefont{H.}~\bibnamefont{Yu}},
  \bibinfo{author}{\bibfnamefont{B.}~\bibnamefont{Liu}},
  \bibinfo{author}{\bibfnamefont{W.}~\bibnamefont{Tian}}, \bibnamefont{and}
  \bibinfo{author}{\bibfnamefont{T.}~\bibnamefont{Cui}}, \bibinfo{journal}{Sci.
  Rep.} \textbf{\bibinfo{volume}{4}}, \bibinfo{pages}{6968}
  (\bibinfo{year}{2014}{\natexlab{a}}).

\bibitem[{\citenamefont{Duan et~al.}(2014{\natexlab{b}})\citenamefont{Duan,
  Huang, Tian, Li, Yu, Liu, Ma, Liu, and Cui}}]{Duan2014B}
\bibinfo{author}{\bibfnamefont{D.}~\bibnamefont{Duan}},
  \bibinfo{author}{\bibfnamefont{X.}~\bibnamefont{Huang}},
  \bibinfo{author}{\bibfnamefont{F.}~\bibnamefont{Tian}},
  \bibinfo{author}{\bibfnamefont{D.}~\bibnamefont{Li}},
  \bibinfo{author}{\bibfnamefont{H.}~\bibnamefont{Yu}},
  \bibinfo{author}{\bibfnamefont{Y.}~\bibnamefont{Liu}},
  \bibinfo{author}{\bibfnamefont{Y.}~\bibnamefont{Ma}},
  \bibinfo{author}{\bibfnamefont{B.}~\bibnamefont{Liu}}, \bibnamefont{and}
  \bibinfo{author}{\bibfnamefont{T.}~\bibnamefont{Cui}},
  \bibinfo{journal}{arXiv:1501.01784}  (\bibinfo{year}{2014}{\natexlab{b}}).

\bibitem[{\citenamefont{Li et~al.}(2014)\citenamefont{Li, Hao, Liu, Li, and
  Ma}}]{Li2014A}
\bibinfo{author}{\bibfnamefont{Y.}~\bibnamefont{Li}},
  \bibinfo{author}{\bibfnamefont{J.}~\bibnamefont{Hao}},
  \bibinfo{author}{\bibfnamefont{H.}~\bibnamefont{Liu}},
  \bibinfo{author}{\bibfnamefont{Y.}~\bibnamefont{Li}}, \bibnamefont{and}
  \bibinfo{author}{\bibfnamefont{Y.}~\bibnamefont{Ma}}, \bibinfo{journal}{J.
  Chem. Phys.} \textbf{\bibinfo{volume}{140}}, \bibinfo{pages}{174712}
  (\bibinfo{year}{2014}).

\bibitem[{\citenamefont{Strobel et~al.}(2011)\citenamefont{Strobel, Ganesh,
  Somayazulu, Kent, and Hemley}}]{Strobel2011A}
\bibinfo{author}{\bibfnamefont{T.~A.} \bibnamefont{Strobel}},
  \bibinfo{author}{\bibfnamefont{P.}~\bibnamefont{Ganesh}},
  \bibinfo{author}{\bibfnamefont{M.}~\bibnamefont{Somayazulu}},
  \bibinfo{author}{\bibfnamefont{P.~R.~C.} \bibnamefont{Kent}},
  \bibnamefont{and} \bibinfo{author}{\bibfnamefont{R.~J.}
  \bibnamefont{Hemley}}, \bibinfo{journal}{Phys. Rev. Lett.}
  \textbf{\bibinfo{volume}{107}}, \bibinfo{pages}{255503}
  (\bibinfo{year}{2011}).

\bibitem[{\citenamefont{Wang et~al.}(2012)\citenamefont{Wang, Tse, Tanaka,
  Iitaka, and Ma}}]{Wang2012A}
\bibinfo{author}{\bibfnamefont{H.}~\bibnamefont{Wang}},
  \bibinfo{author}{\bibfnamefont{J.~S.} \bibnamefont{Tse}},
  \bibinfo{author}{\bibfnamefont{K.}~\bibnamefont{Tanaka}},
  \bibinfo{author}{\bibfnamefont{T.}~\bibnamefont{Iitaka}}, \bibnamefont{and}
  \bibinfo{author}{\bibfnamefont{Y.}~\bibnamefont{Ma}}, \bibinfo{journal}{Proc.
  Natl. Acad. Sci. USA} \textbf{\bibinfo{volume}{109}}, \bibinfo{pages}{6463}
  (\bibinfo{year}{2012}).

\bibitem[{\citenamefont{Szcz{\c{e}}{\'s}niak and
  Durajski}(2013{\natexlab{a}})}]{Szczesniak2013H}
\bibinfo{author}{\bibfnamefont{R.}~\bibnamefont{Szcz{\c{e}}{\'s}niak}}
  \bibnamefont{and} \bibinfo{author}{\bibfnamefont{A.~P.}
  \bibnamefont{Durajski}}, \bibinfo{journal}{Solid State Sciences}
  \textbf{\bibinfo{volume}{25}}, \bibinfo{pages}{45}
  (\bibinfo{year}{2013}{\natexlab{a}}).

\bibitem[{\citenamefont{Jin et~al.}(2010)\citenamefont{Jin, Meng, He, Ma, Liu,
  Cui, Zou, and Mao}}]{Jin2010A}
\bibinfo{author}{\bibfnamefont{X.}~\bibnamefont{Jin}},
  \bibinfo{author}{\bibfnamefont{X.}~\bibnamefont{Meng}},
  \bibinfo{author}{\bibfnamefont{Z.}~\bibnamefont{He}},
  \bibinfo{author}{\bibfnamefont{Y.}~\bibnamefont{Ma}},
  \bibinfo{author}{\bibfnamefont{B.}~\bibnamefont{Liu}},
  \bibinfo{author}{\bibfnamefont{T.}~\bibnamefont{Cui}},
  \bibinfo{author}{\bibfnamefont{G.}~\bibnamefont{Zou}}, \bibnamefont{and}
  \bibinfo{author}{\bibfnamefont{H.}~\bibnamefont{Mao}},
  \bibinfo{journal}{PNAS} \textbf{\bibinfo{volume}{107}}, \bibinfo{pages}{9969}
  (\bibinfo{year}{2010}).

\bibitem[{\citenamefont{Flores-Livas et~al.}(2012)\citenamefont{Flores-Livas,
  Amsler, Lenosky, Lehtovaara, Botti, Marques, and
  Goedecker}}]{FloresLivas2012A}
\bibinfo{author}{\bibfnamefont{J.~A.} \bibnamefont{Flores-Livas}},
  \bibinfo{author}{\bibfnamefont{M.}~\bibnamefont{Amsler}},
  \bibinfo{author}{\bibfnamefont{T.~J.} \bibnamefont{Lenosky}},
  \bibinfo{author}{\bibfnamefont{L.}~\bibnamefont{Lehtovaara}},
  \bibinfo{author}{\bibfnamefont{S.}~\bibnamefont{Botti}},
  \bibinfo{author}{\bibfnamefont{M.~A.~L.} \bibnamefont{Marques}},
  \bibnamefont{and}
  \bibinfo{author}{\bibfnamefont{S.}~\bibnamefont{Goedecker}},
  \bibinfo{journal}{Phys. Rev. Lett.} \textbf{\bibinfo{volume}{108}},
  \bibinfo{pages}{117004} (\bibinfo{year}{2012}).

\bibitem[{\citenamefont{Szcz{\c{e}}{\'s}niak and
  Durajski}(2013{\natexlab{b}})}]{Szczesniak2013C}
\bibinfo{author}{\bibfnamefont{R.}~\bibnamefont{Szcz{\c{e}}{\'s}niak}}
  \bibnamefont{and} \bibinfo{author}{\bibfnamefont{A.~P.}
  \bibnamefont{Durajski}}, \bibinfo{journal}{J. Phys. Chem. Solids}
  \textbf{\bibinfo{volume}{74}}, \bibinfo{pages}{641}
  (\bibinfo{year}{2013}{\natexlab{b}}).

\bibitem[{\citenamefont{Abe and Ashcroft}(2011)}]{Kazutaka2011A}
\bibinfo{author}{\bibfnamefont{K.}~\bibnamefont{Abe}} \bibnamefont{and}
  \bibinfo{author}{\bibfnamefont{N.~W.} \bibnamefont{Ashcroft}},
  \bibinfo{journal}{Phys. Rev. B} \textbf{\bibinfo{volume}{84}},
  \bibinfo{pages}{104118} (\bibinfo{year}{2011}).

\bibitem[{\citenamefont{Szcz{\c{e}}{\'s}niak
  et~al.}(2013)\citenamefont{Szcz{\c{e}}{\'s}niak, Drzazga, and
  Duda}}]{Szczesniak2013F}
\bibinfo{author}{\bibfnamefont{R.}~\bibnamefont{Szcz{\c{e}}{\'s}niak}},
  \bibinfo{author}{\bibfnamefont{E.~A.} \bibnamefont{Drzazga}},
  \bibnamefont{and} \bibinfo{author}{\bibfnamefont{A.~M.} \bibnamefont{Duda}},
  \bibinfo{journal}{Solid State Commun.} \textbf{\bibinfo{volume}{166}},
  \bibinfo{pages}{50} (\bibinfo{year}{2013}).

\bibitem[{\citenamefont{Li et~al.}(2010)\citenamefont{Li, Gao, Xie, Ma, Cui,
  and Zou}}]{YLiPNAS}
\bibinfo{author}{\bibfnamefont{Y.}~\bibnamefont{Li}},
  \bibinfo{author}{\bibfnamefont{G.}~\bibnamefont{Gao}},
  \bibinfo{author}{\bibfnamefont{Y.}~\bibnamefont{Xie}},
  \bibinfo{author}{\bibfnamefont{Y.}~\bibnamefont{Ma}},
  \bibinfo{author}{\bibfnamefont{T.}~\bibnamefont{Cui}}, \bibnamefont{and}
  \bibinfo{author}{\bibfnamefont{G.}~\bibnamefont{Zou}},
  \bibinfo{journal}{Proc. Natl. Acad. Sci. USA} \textbf{\bibinfo{volume}{107}},
  \bibinfo{pages}{15708} (\bibinfo{year}{2010}).

\bibitem[{\citenamefont{Ikeda}(2001)}]{Ikeda2001A}
\bibinfo{author}{\bibfnamefont{T.}~\bibnamefont{Ikeda}},
  \bibinfo{journal}{Phys. Rev. B} \textbf{\bibinfo{volume}{64}},
  \bibinfo{pages}{104103} (\bibinfo{year}{2001}).

\bibitem[{\citenamefont{Cockcroft and Fitch}(1990)}]{Cockcroft1990A}
\bibinfo{author}{\bibfnamefont{J.~K.} \bibnamefont{Cockcroft}}
  \bibnamefont{and} \bibinfo{author}{\bibfnamefont{A.~N.} \bibnamefont{Fitch}},
  \bibinfo{journal}{Z. Kristallogr.} \textbf{\bibinfo{volume}{193}},
  \bibinfo{pages}{1} (\bibinfo{year}{1990}).

\bibitem[{\citenamefont{Shimizu et~al.}(1991)\citenamefont{Shimizu, Nakamichi,
  and Sasaki}}]{Shimizu1991A}
\bibinfo{author}{\bibfnamefont{H.}~\bibnamefont{Shimizu}},
  \bibinfo{author}{\bibfnamefont{Y.}~\bibnamefont{Nakamichi}},
  \bibnamefont{and} \bibinfo{author}{\bibfnamefont{S.}~\bibnamefont{Sasaki}},
  \bibinfo{journal}{J. Chem. Phys.} \textbf{\bibinfo{volume}{95}},
  \bibinfo{pages}{2036} (\bibinfo{year}{1991}).

\bibitem[{\citenamefont{Shimizu et~al.}(1992)\citenamefont{Shimizu, Murashima,
  and Sasaki}}]{Shimizu1992A}
\bibinfo{author}{\bibfnamefont{H.}~\bibnamefont{Shimizu}},
  \bibinfo{author}{\bibfnamefont{H.}~\bibnamefont{Murashima}},
  \bibnamefont{and} \bibinfo{author}{\bibfnamefont{S.}~\bibnamefont{Sasaki}},
  \bibinfo{journal}{J. Chem. Phys.} \textbf{\bibinfo{volume}{97}},
  \bibinfo{pages}{7137} (\bibinfo{year}{1992}).

\bibitem[{\citenamefont{Endo et~al.}(1994)\citenamefont{Endo, Ichimiya, Koto,
  Sasaki, and Shimizu}}]{Endo1994A}
\bibinfo{author}{\bibfnamefont{S.}~\bibnamefont{Endo}},
  \bibinfo{author}{\bibfnamefont{N.}~\bibnamefont{Ichimiya}},
  \bibinfo{author}{\bibfnamefont{K.}~\bibnamefont{Koto}},
  \bibinfo{author}{\bibfnamefont{S.}~\bibnamefont{Sasaki}}, \bibnamefont{and}
  \bibinfo{author}{\bibfnamefont{H.}~\bibnamefont{Shimizu}},
  \bibinfo{journal}{Phys. Rev. B} \textbf{\bibinfo{volume}{50}},
  \bibinfo{pages}{5865} (\bibinfo{year}{1994}).

\bibitem[{\citenamefont{Endo et~al.}(1996)\citenamefont{Endo, Honda, Sasaki,
  Shimizu, Shimomura, and Kikegawa}}]{Endo1996A}
\bibinfo{author}{\bibfnamefont{S.}~\bibnamefont{Endo}},
  \bibinfo{author}{\bibfnamefont{A.}~\bibnamefont{Honda}},
  \bibinfo{author}{\bibfnamefont{S.}~\bibnamefont{Sasaki}},
  \bibinfo{author}{\bibfnamefont{H.}~\bibnamefont{Shimizu}},
  \bibinfo{author}{\bibfnamefont{O.}~\bibnamefont{Shimomura}},
  \bibnamefont{and} \bibinfo{author}{\bibfnamefont{T.}~\bibnamefont{Kikegawa}},
  \bibinfo{journal}{Phys. Rev. B} \textbf{\bibinfo{volume}{54}},
  \bibinfo{pages}{R717} (\bibinfo{year}{1996}).

\bibitem[{\citenamefont{Fujihisa et~al.}(1998)\citenamefont{Fujihisa, Yamawaki,
  Sakashita, Aoki, Sasaki, and Shimizu}}]{Fujihisa1998A}
\bibinfo{author}{\bibfnamefont{H.}~\bibnamefont{Fujihisa}},
  \bibinfo{author}{\bibfnamefont{H.}~\bibnamefont{Yamawaki}},
  \bibinfo{author}{\bibfnamefont{M.}~\bibnamefont{Sakashita}},
  \bibinfo{author}{\bibfnamefont{K.}~\bibnamefont{Aoki}},
  \bibinfo{author}{\bibfnamefont{S.}~\bibnamefont{Sasaki}}, \bibnamefont{and}
  \bibinfo{author}{\bibfnamefont{H.}~\bibnamefont{Shimizu}},
  \bibinfo{journal}{Phys. Rev. B} \textbf{\bibinfo{volume}{57}},
  \bibinfo{pages}{2651} (\bibinfo{year}{1998}).

\bibitem[{\citenamefont{Sakashita et~al.}(1997)\citenamefont{Sakashita,
  Yamawaki, Fujihisa, Aoki, Sasaki, and Shimizu}}]{Sakashita1997A}
\bibinfo{author}{\bibfnamefont{M.}~\bibnamefont{Sakashita}},
  \bibinfo{author}{\bibfnamefont{H.}~\bibnamefont{Yamawaki}},
  \bibinfo{author}{\bibfnamefont{H.}~\bibnamefont{Fujihisa}},
  \bibinfo{author}{\bibfnamefont{K.}~\bibnamefont{Aoki}},
  \bibinfo{author}{\bibfnamefont{S.}~\bibnamefont{Sasaki}}, \bibnamefont{and}
  \bibinfo{author}{\bibfnamefont{H.}~\bibnamefont{Shimizu}},
  \bibinfo{journal}{Phys. Rev. Lett.} \textbf{\bibinfo{volume}{79}},
  \bibinfo{pages}{1082} (\bibinfo{year}{1997}).

\bibitem[{\citenamefont{Fujihisa et~al.}(2004)\citenamefont{Fujihisa, Yamawaki,
  Sakashita, Nakayama, Yamada, and Aoki}}]{Fujihisa2004A}
\bibinfo{author}{\bibfnamefont{H.}~\bibnamefont{Fujihisa}},
  \bibinfo{author}{\bibfnamefont{H.}~\bibnamefont{Yamawaki}},
  \bibinfo{author}{\bibfnamefont{M.}~\bibnamefont{Sakashita}},
  \bibinfo{author}{\bibfnamefont{A.}~\bibnamefont{Nakayama}},
  \bibinfo{author}{\bibfnamefont{T.}~\bibnamefont{Yamada}}, \bibnamefont{and}
  \bibinfo{author}{\bibfnamefont{K.}~\bibnamefont{Aoki}},
  \bibinfo{journal}{Phys. Rev. B} \textbf{\bibinfo{volume}{69}},
  \bibinfo{pages}{214102} (\bibinfo{year}{2004}).

\bibitem[{\citenamefont{Shimizu et~al.}(1997)\citenamefont{Shimizu, Ushida,
  Sasaki, Sakashita, Yamawaki, and Aoki}}]{Shimizu1997A}
\bibinfo{author}{\bibfnamefont{H.}~\bibnamefont{Shimizu}},
  \bibinfo{author}{\bibfnamefont{T.}~\bibnamefont{Ushida}},
  \bibinfo{author}{\bibfnamefont{S.}~\bibnamefont{Sasaki}},
  \bibinfo{author}{\bibfnamefont{M.}~\bibnamefont{Sakashita}},
  \bibinfo{author}{\bibfnamefont{H.}~\bibnamefont{Yamawaki}}, \bibnamefont{and}
  \bibinfo{author}{\bibfnamefont{K.}~\bibnamefont{Aoki}},
  \bibinfo{journal}{Phys. Rev. B} \textbf{\bibinfo{volume}{55}},
  \bibinfo{pages}{5538} (\bibinfo{year}{1997}).

\bibitem[{\citenamefont{Marsiglio et~al.}(1988)\citenamefont{Marsiglio,
  Schossmann, and Carbotte}}]{Marsiglio1988A}
\bibinfo{author}{\bibfnamefont{F.}~\bibnamefont{Marsiglio}},
  \bibinfo{author}{\bibfnamefont{M.}~\bibnamefont{Schossmann}},
  \bibnamefont{and} \bibinfo{author}{\bibfnamefont{J.~P.}
  \bibnamefont{Carbotte}}, \bibinfo{journal}{Phys. Rev. B}
  \textbf{\bibinfo{volume}{37}}, \bibinfo{pages}{4965} (\bibinfo{year}{1988}).

\bibitem[{\citenamefont{Eliashberg}(1960)}]{Eliashberg1960A}
\bibinfo{author}{\bibfnamefont{G.~M.} \bibnamefont{Eliashberg}},
  \bibinfo{journal}{Soviet Physics JETP} \textbf{\bibinfo{volume}{11}},
  \bibinfo{pages}{696} (\bibinfo{year}{1960}).

\bibitem[{\citenamefont{Morel and Anderson}(1962)}]{Morel1962A}
\bibinfo{author}{\bibfnamefont{P.}~\bibnamefont{Morel}} \bibnamefont{and}
  \bibinfo{author}{\bibfnamefont{P.~W.} \bibnamefont{Anderson}},
  \bibinfo{journal}{Phys. Rev.} \textbf{\bibinfo{volume}{125}},
  \bibinfo{pages}{1962} (\bibinfo{year}{1962}).

\bibitem[{\citenamefont{Szcz{\c{e}}{\'s}niak and
  Durajski}(2014{\natexlab{a}})}]{Szczesniak2014B}
\bibinfo{author}{\bibfnamefont{R.}~\bibnamefont{Szcz{\c{e}}{\'s}niak}}
  \bibnamefont{and} \bibinfo{author}{\bibfnamefont{A.~P.}
  \bibnamefont{Durajski}}, \bibinfo{journal}{Superconductor Science and
  Technology} \textbf{\bibinfo{volume}{27}}, \bibinfo{pages}{015003}
  (\bibinfo{year}{2014}{\natexlab{a}}).

\bibitem[{\citenamefont{Durajski and
  Szcz{\c{e}}{\'s}niak}(2014)}]{Szczesniak2014C}
\bibinfo{author}{\bibfnamefont{A.~P.} \bibnamefont{Durajski}} \bibnamefont{and}
  \bibinfo{author}{\bibfnamefont{R.}~\bibnamefont{Szcz{\c{e}}{\'s}niak}},
  \bibinfo{journal}{Superconductor Science and Technology}
  \textbf{\bibinfo{volume}{27}}, \bibinfo{pages}{115012}
  (\bibinfo{year}{2014}).

\bibitem[{\citenamefont{Szcz{\c{e}}{\'s}niak and
  Durajski}(2014{\natexlab{b}})}]{Szczesniak2014E}
\bibinfo{author}{\bibfnamefont{R.}~\bibnamefont{Szcz{\c{e}}{\'s}niak}}
  \bibnamefont{and} \bibinfo{author}{\bibfnamefont{A.~P.}
  \bibnamefont{Durajski}}, \bibinfo{journal}{Superconductor Science and
  Technology} \textbf{\bibinfo{volume}{27}}, \bibinfo{pages}{125004}
  (\bibinfo{year}{2014}{\natexlab{b}}).

\bibitem[{\citenamefont{Eschrig}(2001)}]{Eschrig2001A}
\bibinfo{author}{\bibfnamefont{H.}~\bibnamefont{Eschrig}},
  \emph{\bibinfo{title}{Theory of superconductivity a primer}}
  (\bibinfo{publisher}{Citeseer}, \bibinfo{year}{2001}).

\bibitem[{\citenamefont{Varelogiannis}(1997)}]{Varelogiannis1997A}
\bibinfo{author}{\bibfnamefont{G.}~\bibnamefont{Varelogiannis}},
  \bibinfo{journal}{Z. Phys. B} \textbf{\bibinfo{volume}{104}},
  \bibinfo{pages}{411} (\bibinfo{year}{1997}).

\bibitem[{\citenamefont{Allen and Dynes}(1975)}]{Allen1975A}
\bibinfo{author}{\bibfnamefont{P.~B.} \bibnamefont{Allen}} \bibnamefont{and}
  \bibinfo{author}{\bibfnamefont{R.~C.} \bibnamefont{Dynes}},
  \bibinfo{journal}{Phys. Rev. B} \textbf{\bibinfo{volume}{905}},
  \bibinfo{pages}{1975} (\bibinfo{year}{1975}).

\bibitem[{\citenamefont{McMillan}(1968)}]{McMillan1968A}
\bibinfo{author}{\bibfnamefont{W.~L.} \bibnamefont{McMillan}},
  \bibinfo{journal}{Phys. Rev.} \textbf{\bibinfo{volume}{167}},
  \bibinfo{pages}{331} (\bibinfo{year}{1968}).

\bibitem[{\citenamefont{Durajski et~al.}(2012)\citenamefont{Durajski,
  Szcz{\c{e}}{\'s}niak, and Jarosik}}]{Szczesniak2012K}
\bibinfo{author}{\bibfnamefont{A.~P.} \bibnamefont{Durajski}},
  \bibinfo{author}{\bibfnamefont{R.}~\bibnamefont{Szcz{\c{e}}{\'s}niak}},
  \bibnamefont{and} \bibinfo{author}{\bibfnamefont{M.~W.}
  \bibnamefont{Jarosik}}, \bibinfo{journal}{Phase Transitions}
  \textbf{\bibinfo{volume}{85}}, \bibinfo{pages}{727} (\bibinfo{year}{2012}).

\bibitem[{\citenamefont{Szcz{\c{e}}{\'s}niak and
  Jarosik}(2012)}]{Szczesniak2012L}
\bibinfo{author}{\bibfnamefont{R.}~\bibnamefont{Szcz{\c{e}}{\'s}niak}}
  \bibnamefont{and} \bibinfo{author}{\bibfnamefont{M.~W.}
  \bibnamefont{Jarosik}}, \bibinfo{journal}{Acta Phys. Pol. A}
  \textbf{\bibinfo{volume}{121}}, \bibinfo{pages}{841} (\bibinfo{year}{2012}).

\bibitem[{\citenamefont{Bardeen
  et~al.}(1957{\natexlab{a}})\citenamefont{Bardeen, Cooper, and
  Schrieffer}}]{Bardeen1957A}
\bibinfo{author}{\bibfnamefont{J.}~\bibnamefont{Bardeen}},
  \bibinfo{author}{\bibfnamefont{L.~N.} \bibnamefont{Cooper}},
  \bibnamefont{and} \bibinfo{author}{\bibfnamefont{J.~R.}
  \bibnamefont{Schrieffer}}, \bibinfo{journal}{Phys. Rev.}
  \textbf{\bibinfo{volume}{106}}, \bibinfo{pages}{162}
  (\bibinfo{year}{1957}{\natexlab{a}}).

\bibitem[{\citenamefont{Bardeen
  et~al.}(1957{\natexlab{b}})\citenamefont{Bardeen, Cooper, and
  Schrieffer}}]{Bardeen1957B}
\bibinfo{author}{\bibfnamefont{J.}~\bibnamefont{Bardeen}},
  \bibinfo{author}{\bibfnamefont{L.~N.} \bibnamefont{Cooper}},
  \bibnamefont{and} \bibinfo{author}{\bibfnamefont{J.~R.}
  \bibnamefont{Schrieffer}}, \bibinfo{journal}{Phys. Rev.}
  \textbf{\bibinfo{volume}{108}}, \bibinfo{pages}{1175}
  (\bibinfo{year}{1957}{\natexlab{b}}).

\bibitem[{\citenamefont{Carbotte}(1990)}]{Carbotte1990A}
\bibinfo{author}{\bibfnamefont{J.}~\bibnamefont{Carbotte}},
  \bibinfo{journal}{Rev. Mod. Phys.} \textbf{\bibinfo{volume}{62}},
  \bibinfo{pages}{1027} (\bibinfo{year}{1990}).

\end{thebibliography}

%
%
\end{document}